\documentclass[sigconf]{acmart}

\AtBeginDocument{
  \providecommand\BibTeX{{
    \normalfont B\kern-0.5em{\scshape i\kern-0.25em b}\kern-0.8em\TeX}}}

\copyrightyear{2022}
\acmYear{2022}
\setcopyright{acmcopyright}\acmConference[DEBS '22]{The 16th ACM International Conference on Distributed and Event-based Systems}{June 27--30, 2022}{Copenhagen, Denmark}
\acmBooktitle{The 16th ACM International Conference on Distributed and Event-based Systems (DEBS '22), June 27--30, 2022, Copenhagen, Denmark}
\acmPrice{15.00}
\acmDOI{10.1145/3524860.3539639}
\acmISBN{978-1-4503-9308-9/22/06}

\usepackage{multirow}
\usepackage{cleveref}
\usepackage{hhline}
\usepackage[font=small,labelfont=bf]{caption}

\newcommand{\first}{\emph{(i)} } 
\newcommand{\ii}{\emph{(ii)} }
\newcommand{\iii}{\emph{(iii)} }
\newcommand{\iv}{\emph{(iv)} }
\newcommand{\feature}[1]{\texttt{\small{#1}}}

\newcommand*\circles[1]{\raisebox{.5pt}{\textcircled{\raisebox{-.9pt} {#1}}}}

\begin{document}
\title{Zero-Shot Cost Models for Distributed Stream Processing}
\settopmatter{authorsperrow=4}

\author{Roman Heinrich}
\affiliation{
  \institution{DHBW Mannheim}
  \city{}
  \country{}
}

\author{Manisha Luthra}
\affiliation{
  \institution{Technical University\\of Darmstadt}
  \city{}
  \country{}
}

\author{Harald Kornmayer}
\affiliation{
  \institution{DHBW Mannheim}
  \city{}
  \country{}
}

\author{Carsten Binnig}
\affiliation{
  \institution{Technical University of Darmstadt \& DFKI}
  \city{}
  \country{}
}

\begin{abstract}
This paper proposes a learned cost estimation model for Distributed Stream Processing Systems (DSPS) with an aim to provide accurate cost predictions of executing queries. 
A major premise of this work is that the proposed learned model can generalize to the dynamics of streaming workloads \emph{out-of-the-box}.
This means a model once trained can accurately predict performance metrics such as  \emph{latency} and \emph{throughput} even if the characteristics of the data and workload or the deployment of operators to hardware changes at runtime.
That way, the model can be used to solve tasks such as optimizing the placement of operators to minimize the end-to-end latency of a streaming query or maximize its throughput even under varying conditions. 
Our evaluation on a well-known DSPS, Apache Storm, shows that the model can predict accurately for unseen workloads and queries while generalizing across real-world benchmarks. 
\end{abstract}



\maketitle
\section{Introduction}

\begin{figure}
\centering
\includegraphics[width=0.7\linewidth]{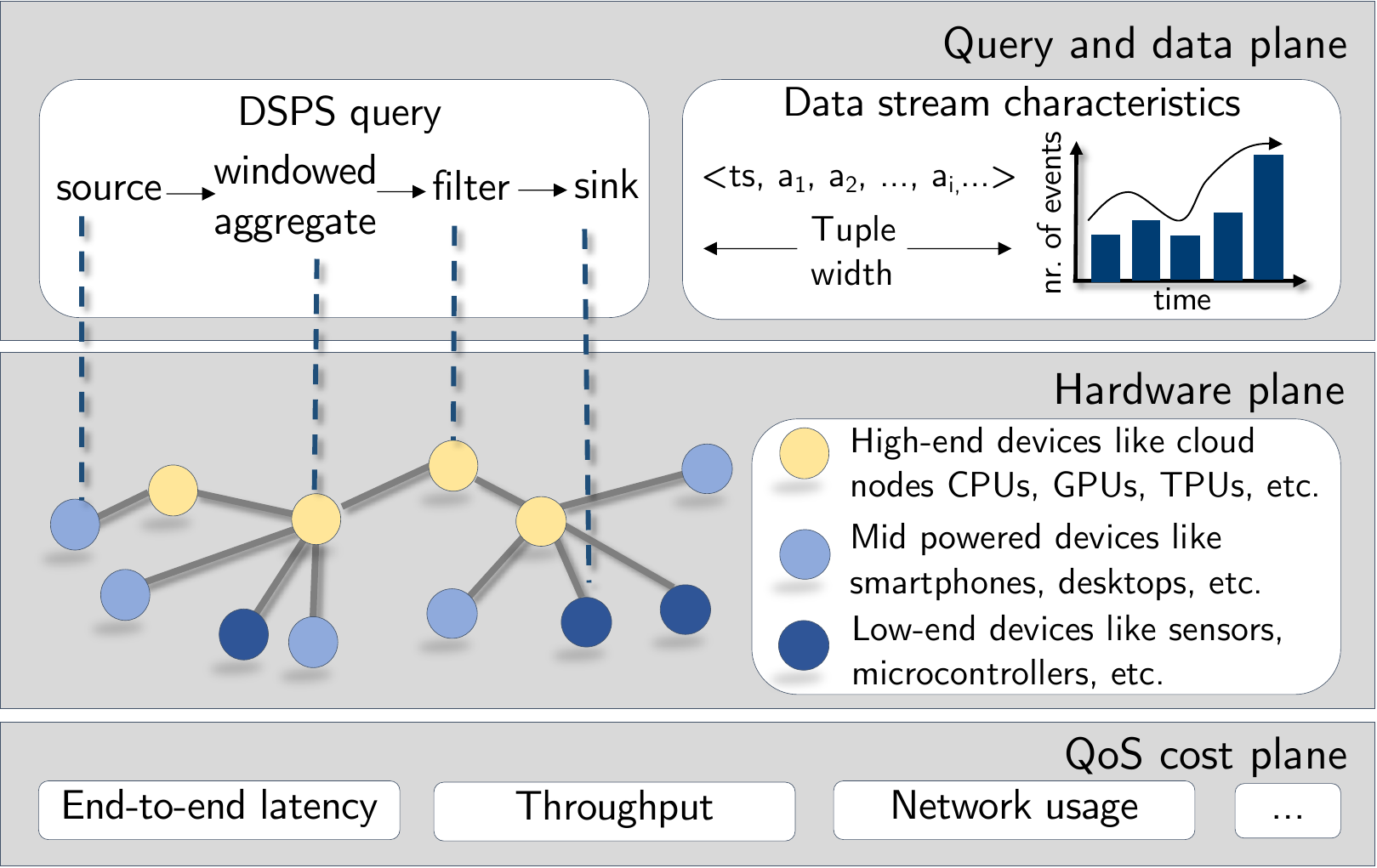}
\vspace{-2.5ex}
\caption{A DSPS has to provide guarantees in terms of one or more quality-of-service (QoS) cost metrics such as latency and throughput. The challenge is that DSPS serve a wide range of workloads on potentially diverse hardware, which makes the cost estimation harder.}
\label{fig:parameter_space}
\vspace{-4.5ex}
\end{figure}

\paragraph{Motivation}
Distributed Stream Processing System (DSPS) correlates and analyzes data streams from multiple data sources to derive higher-level information for a wide range of applications. 
At the core, a DSPS takes a continuous query that represents one or more tasks for the given application and processes the query in a distributed way using multiple hardware resources (cf. \Cref{fig:parameter_space}).
While doing so, for many applications, a DSPS has to provide guarantees in terms of one or more quality-of-service (QoS) cost metrics such as latency and throughput.
For instance, in Facebook, queries for click stream analytics have to provide a very high throughput to process input event streams at around 9 GiB/s and at the same time also ensure low latency \cite{shao2011}.

Typically, a DSPS provides QoS guarantees using optimization mechanisms such as \emph{operator placement} that usually monitors the costs to decide on the mapping of operators to hardware as shown in \Cref{fig:parameter_space} \cite{nardelli2019}.
Moreover, frequent reconfigurations of the operator placement are required based on the observed changes of the workload (e.g., during peaks in the input stream).
Further, to provide QoS guarantees, a DSPS uses multiple other techniques such as \emph{elasticity} or adaption of \emph{operator parallelism}~\cite{russo2019}. ~\cite{hirzel2014} provides an extensive list of other DSPS optimization mechanisms.

However, many of these optimization mechanisms classically rely on heuristics~\cite{nardelli2019,eskandari2021} or other analytical approaches like queueing theory~\cite{mencagli2018} that approximates the effects of placement decisions, the degree of parallelism, etc., on the actual latency and throughput. 
As a result, these techniques often {make simplifying assumptions on the estimation 
of the required QoS and thus take non-optimal decisions (e.g., they decide on a non-optimal placement). 

More recently, machine learning (ML) has been used in solving these optimization tasks such as to place operators~\cite{luthra2021,li2018}, to auto-scale DSPS~\cite{russo2019} or even to estimate resource utilization~\cite{wang2017}, 
which has shown promising initial results~\cite{ahmad2021}.
However, the existing ML-based approaches cannot be used out-of-the-box as these often are specialized either for the given optimization task (e.g., elasticity) or support only a restricted set of workloads and hardware resources.

While, in this paper, we argue that learned cost models should be used for DSPS to allow better decisions for a broad set of optimization tasks.
For example, the cost model can be used in advance before actually placing operators to different resources to know the expected throughput and latency.
The same observations hold for other optimization tasks, such as determining the degree of parallelism for an operator.
Recently, learned cost models have also been explored in database systems to estimate costs such as query runtime and thus solve problems related to query optimization or scheduling. 
Yet these learned models are highly \emph{workload-driven}~\cite{li2021} -- meaning that they make strong assumptions about what data and queries as well as hardware should be supported.


Thus, a core problem of existing learned cost models is that they can hardly generalize to the changes in the workload (queries and data) or even the placement on different hardware platforms at runtime.
Consequently, either a separate model must be trained for a combination of workload and hardware or the existing one must be constantly retrained, which not only causes high overhead but also does not allow the DSPS to react instantly to changes.

\vspace{-2ex}\paragraph{Contributions}
In this paper, therefore, we propose a new approach for \emph{learned cost models} for DSPSs based on the concept of \emph{zero-shot learning}~\cite{hilprecht2022, hilprecht2022cidr}.
The key idea is to train a model on a broad spectrum of different streaming workloads and hardware resources to enable generalization.
As such, a zero-shot cost model for DSPS can provide accurate estimates even under the changes in the streaming workload or for different hardware platforms that are used for placing operators.
Moreover, the zero-shot cost model can even adapt to an entirely \emph{new} workload \emph{out-of-the-box}, with queries and data characteristics the model has not seen during the training.
Thus, our approach comes as a huge benefit for DSPSs since they are known to be highly dynamic in nature (e.g., fluctuations in input stream).
That way, using such a zero-shot cost model, several DSPS optimization tasks can be performed (e.g., operator placement) in a highly accurate manner such that the desired QoS guarantees are satisfied even under dynamics in streaming workload.

A key to enable a zero-shot cost model is a novel model architecture that represents data, queries and hardware as input for our cost model.
At the core of our model architecture is a so-called \emph{transferable feature representation} that allows the learned cost model to generalize to a broader set of workloads or even be used across hardware platforms.
For instance, to make predictions under changing workloads, the transferable feature representation relies on general information such as \emph{event rate at the source} and other information such as \emph{tuple width} of data streams or \emph{window size} of operators.
Another important aspect of the transferable feature representation is that we include so-called \emph{data characteristics} of the queries and data streams as features, such as selectivity of an operator, so that the zero-shot model can learn the runtime behavior of a query.

In summary, this work makes the following contributions: \newline
\first We discuss the training and inference procedure of the zero-shot model. For training, we provide a broad spectrum of workloads and hardware platforms to a zero-shot cost model for DSPS such that it can learn to generalize to unseen streaming workloads and provide estimates across different hardware platforms. \\
\ii We present a new model architecture for learning such a zero-shot cost model for DSPSs that can generalize to unseen workloads by using a transferable feature representation. \\
\iii We provide an evaluation of our model on Apache Storm, a well-known DSPS. The results show that our model accurately predicts the performance of DSPS queries on Storm, even for unseen operator and data stream parameters and generalizes across different real-world benchmarks without explicitly training for them.

\vspace{-2ex}
\paragraph{Outline}
In \Cref{sec:overview}, we provide an overview of our approach before we explain the details of the model architecture and its feature representation in \Cref{sec:model}.
Afterwards, we present the experiments in \Cref{sec:evaluation}, related work in \Cref{sec:related_work}, and conclude in \Cref{sec:conclusion}.

\vspace{-2ex}
\section{Overview of Approach}
\label{sec:overview}
In the following, we provide a high-level overview of our approach. We first start with the idea of how a zero-shot cost model is being trained for DSPS before we then explain its usage at runtime to estimate the cost for guaranteeing QoS.

\vspace{-1.5ex}
\paragraph{Training a Zero-Shot Cost Model}
The proposed zero-shot cost model learns from previous query executions and their observed costs in a supervised manner.
To allow the model to generalize, we generate and train our model with a broad spectrum of queries and streaming data as well as different observed cost metrics (i.e., latency and throughput) that are induced by these queries.
To be more precise, to enable a high variety in the training data, we represent standard query structures for DSPS and vary them in terms of complexity and operators properties such as \emph{window size} (cf. \Cref{sec:evaluation} to see our training range). 
Similarly, we diversify the input data streams by training for several \emph{event rates}, and we capture many different streaming workloads.
Moreover, during training, operators of queries are also deployed on different hardware platforms.
While this would seem like a huge effort, it is a one-time training effort in contrast to state-of-the-art learned approaches that need to train a model per streaming workload (data and query). 
The main idea to enable generalization across streaming workloads and hardware platforms is our model architecture that relies on the aforementioned transferable feature representation of workloads and hardware (cf. \Cref{sec:model}).

\vspace{-1.5ex}
\paragraph{Using a Zero-Shot Cost Model}
Once a zero-shot model is trained, it can be used at runtime to predict cost metrics for an unseen query across different hardware platforms. 
In particular, our model can infer costs accurately for an unseen query with an entirely different data distribution of input data stream and extrapolate for unseen operator properties, e.g., window size.

Consequently, we envision zero-shot cost models as a foundation for complex optimization tasks like the operator placement problem. 
For instance, cost predictions of operator placement on different hardware platforms can be used to find a hardware resource with an objective to satisfy a certain latency constraint or demands on throughput.
Moreover, such cost predictions can also serve as a basis for other optimization tasks, such as to determine the right {parallelization degree} or the number of resources for deployment.
While the combination of zero-shot models with these optimization tasks is clearly an interesting direction, in this paper, we focus only on how to enable the cost estimation using zero-shot models.
\vspace{-1.5ex}
\section{Zero-Shot Cost Models for DSPS}
\label{sec:model}

In \Cref{subsec:overall_approach}, we first describe the cost metrics and the proposed model architecture.
Afterwards, we discuss our transferable feature representation in \Cref{subsec:features} and conclude with the training and inference procedure in \Cref{subsec:learning_arch}.

\vspace{-2ex}
\subsection{QoS Metrics and Model Architecture}
\label{subsec:overall_approach}
Our approach draws inspiration from the zero-shot cost models for databases~\cite{hilprecht2022} that aims to predict query runtimes across unseen relational databases. 
However, \cite{hilprecht2022, hilprecht2022cidr} is not easily extensible for DSPS because the characteristics of databases and their queries differ largely from that of a DSPS, as discussed in the following. 

First, different from databases, a query is continuous in a DSPS and is composed of a logical set of \textit{streaming operators} ($\Omega$), i.e., operators that operate on unbounded data streams ($D$) instead of tables.
Streaming queries are composed of multiple operators in a so-called \textit{operator graph} ($G$). 
In this graph, each vertex represents an operator ($\omega \in \Omega$), and the edge between them represents the data stream $D$.
Hence, the operator graph describes the logical flow of data streams from one or multiple \emph{sources} (data producer) to the \emph{sink} (data consumer).
The data stream $D$ that flows through the operator graph represents an unbounded set of tuples.

Other differences from databases are: \first The operator graph in a DSPS contains very different operators, e.g., window operator is typically used to bound the data stream. 
\ii For DSPS, different QoS metrics need to be predicted by a cost model for continuous queries that varies over the time. 
For this paper, we consider \emph{latency} and \emph{throughput} for streaming queries in DSPSs that need to be predicted by our learned cost model.
In the following, we now discuss the definitions for those metrics, which we use in this paper since several different definitions have been used in the literature~\cite{karimov2018, li2018, carbone2015}. Afterwards, we then explain the details of our model architecture.

\vspace{-1.5ex}
\paragraph{QoS Metrics} While throughput is well-defined for DSPS, there is no unique definition of latency~\cite{nardelli2019}. 
The reason is that there are several stages of an event tuple (production, ingestion and processing) at which it is timestamped and thus different classes of latency exist~\cite{carbone2015}. 
In this paper, we only consider the so-called \emph{end-to-end latency} that includes all stages of an event. 

\vspace{-1ex} \begin{definition}
\emph{End-to-end latency}: 
For each output tuple $d_O$, the end-to-end latency is the interval between the time at which the oldest input event tuple $d_I$ involved in producing the output tuple $d_O$ is generated at the source and the time that $d_O$ arrives at the sink. In the paper, we use the average end-to-end latency of all the output tuples that arrive at the sink for a given query. 
\end{definition}

\vspace{-2ex} \begin{definition}
\emph{Throughput}: We define the second metric \emph{throughput} for our cost models inline with the literature ~\cite{karimov2018}.
For the execution of a given query, throughput is the number of output tuples that arrive at the sink per time unit. 
\end{definition}

\begin{figure}
    \centering
    \includegraphics[width=0.9\linewidth]{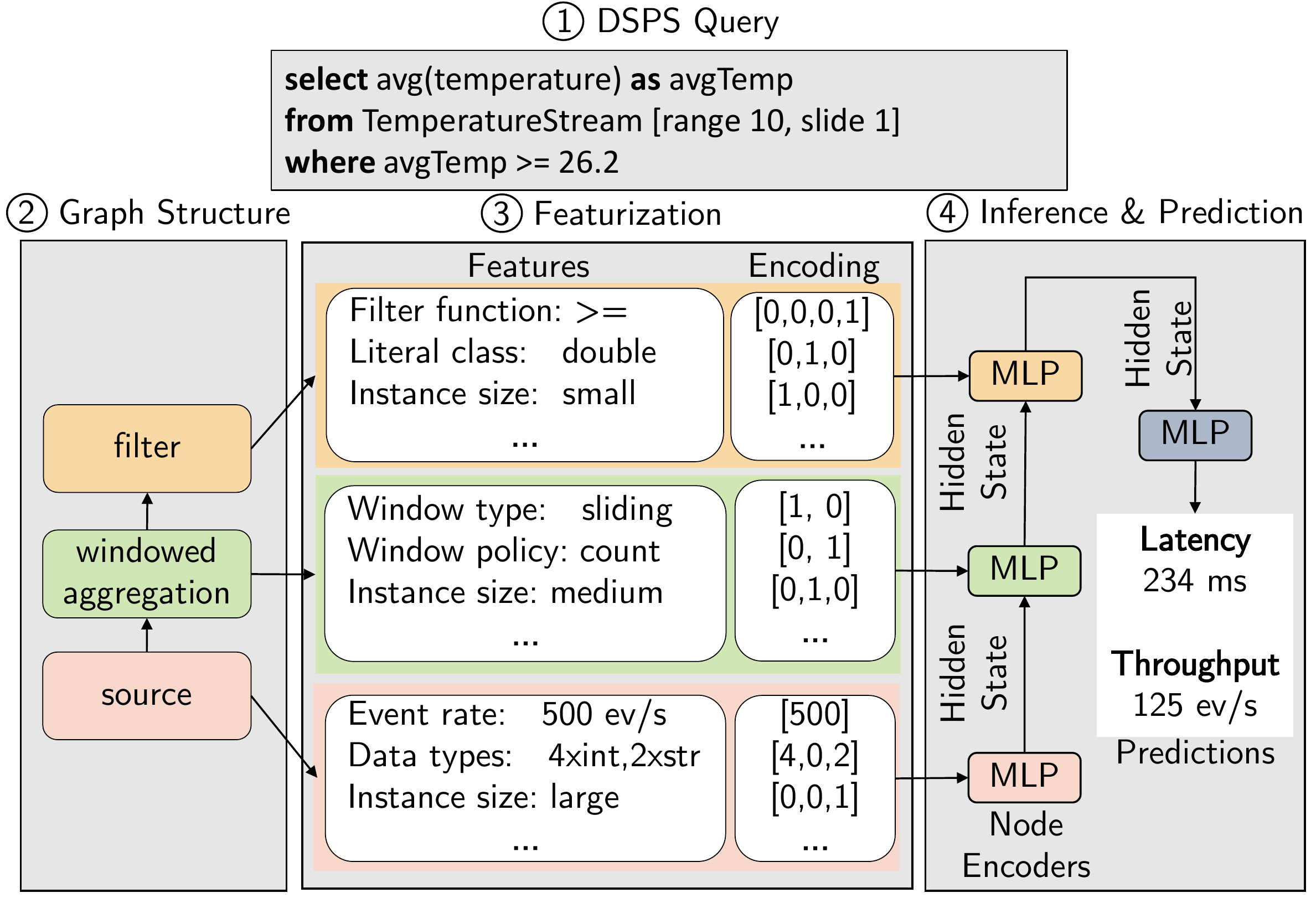}
    \vspace{-3.5ex}
    \caption{Zero-shot cost estimation for DSPS. For predicting costs a streaming query \circles{1} is transformed into a graph structure \circles{2} which uses a novel transferable representation \circles{3}. For cost predictions \circles{4}, we use the graph-based structure where features of every operator are encoded in a separate multi-layer perceptron (MLP) and propagated along the graph using the order of operators in the query. A final MLP predicts throughput and latency.}
    \label{fig:overall_approach}
    \vspace{-3ex}
\end{figure}

\vspace{-2ex}
\paragraph{Model Architecture}
For realizing zero-shot models for DSPS that can predict these QoS metrics, a core question we ask is, \emph{``what can be learned from a given query and data stream that can be generalized for most of the streaming workloads?''}. 
We answer this by using a new model architecture that aims to represent streaming queries using a graph representation with so-called transferable features (\Cref{subsec:features}).
The overall approach of the model architecture is explained by using an example as shown in~\Cref{fig:overall_approach}.
\circles{1} Given a DSPS query that computes whether an average temperature from a data stream \feature{TemperatureStream} exceeds a threshold, 
\circles{2} we represent each operator of the query as a \emph{node} in our graph structure.
For instance, \feature{source}, \feature{windowed aggregation} and \feature{filter} operators are represented as nodes in the model shown in the example.
\circles{3} For each graph node, we use transferable features that describe operator properties (e.g., \feature{filter} predicate), data characteristics (e.g., \feature{selectivity}) and hardware properties (e.g., \feature{instance size}). 
\circles{4} Once the graph representation is constructed, the model can predict the cost (e.g., throughput and end-to-end latency).
The training and inference procedure is explained in detail in \Cref{subsec:learning_arch}.
Overall, the generic representation based on graphs allows us to make predictions for different query structures. 

\vspace{-1.5ex}
\subsection{Transferable Feature Representation} \label{subsec:features}
Query execution costs like latency and throughput depend heavily on the operator parameters, the data characteristics of the streams and the operator placement. To describe these aspects and make them usable for learning, we propose transferable features. For this, we derive \textit{nodes} from the operator graph $G$ (excluding the sink) and collect transferable operator-related,  data-related, and hardware-related input features for each of them.
By \textit{transferable}, we mean that these features can be applied for any arbitrary streaming workload (data and query). This representation targets commonly used operators for DSPS (e.g. filter, join, aggregation, window). A listing of all nodes with the features we use for our cost model is presented in \Cref{tab:features} and explained as follows. 

\begin{table}
\scriptsize
\begin{tabular}{p{0.6cm}p{0.8cm}p{2.2cm}l}
\hline
\textbf{Node} & \textbf{Category} & \textbf{Feature} & \textbf{Description} \\ \hline
\multirow{3}{*}{all} & hardware & \texttt{instance size} & Properties of the hardware \\
 & data & \texttt{tuple width in} & Averaged incoming tuple width \\
 & data & \texttt{tuple width out} & Outgoing tuple width \\ \hline
\multirow{2}{*}{source} & data & \texttt{input event rate} & Event rate emitted by the source \\
 & data & \texttt{tuple data type} & Data type for each value in tuple \\ \hline
\multirow{3}{*}{filter} & operator & \texttt{filter function} & Comparison function  \\
 & operator & \texttt{literal data type} & Data type of comparison literal \\
 & data & \texttt{selectivity} & see \Cref{def:filter_sel} \\ \hline
\multirow{2}{*}{join} & operator & \texttt{join-key data type} & Data type of the join key \\
 & data & \texttt{selectivity} & see \Cref{def:join_sel} \\ \hline
\multirow{3}{*}{agg.} & operator & \texttt{agg. function} & Aggregation function \\
 & operator & \texttt{group-by data type} & Data type of group-by attribute \\
 & operator & \texttt{agg. data type} & Data type of each value to aggregate \\
 & data & \texttt{selectivity} & see \Cref{def:agg_sel} \\ \hline
\multirow{4}{*}{window} & operator & \texttt{window type} & Shifting strategy (sliding/tumbling) \\
 & operator & \texttt{window policy} & Counting mode (count/time-based) \\
 & operator & \texttt{window size} & Size of the window \\
 & operator & \texttt{slide size} & Size of the sliding interval \\ \hline
\end{tabular}
\caption{Transferable features categorized as operator-, data- and hardware-related. The zero-shot model can learn from them and can be applied to different streaming workloads (data and queries).}
\label{tab:features}
\vspace{-10ex}
\end{table}

\vspace{-1ex}
\paragraph{Operator-related features}
The main idea behind operator-related features is to only include properties as features that describe the operator logic and are transferable for \emph{any} query and DSPS.
So, instead of using non-transferable features of operators, such as \emph{literals} of a filter operator (e.g. \feature{tuple temperature = 5)}, which might have a very different meaning for different data streams, we use generic features such as \feature{data type} of literal and the \feature{function} of a filter predicate (e.g., an equality predicate or a range predicate). 
 
\vspace{-1ex}
\paragraph{Data-related features}
A major aspect of zero-shot models is that it is agnostic of the underlying data distribution of the streaming workload.
Thus, instead of using features like \emph{attribute names} to encode the semantics of the data, we learn from data characteristics (DCs) such as \feature{tuple width} and \feature{event rate} that can be transferred to any workload. 
While DCs like \feature{tuple width} are easily derived, other DCs, such as \feature{selectivity} of operators, are harder to derive or are even not well-defined for operators like windowed joins~\cite{cugola2012}. 
In the following, we thus first define the selectivity for the streaming operators shown in \Cref{tab:features}:

\begin{definition}
\label{def:filter_sel} 
\emph{Filter selectivity}:
The selectivity $sel ({\omega_\sigma})$ of a filter operator $\omega_\sigma$ is the ratio of the number of outgoing to incoming tuples in the input stream $D$, formally as:

\vspace{-2ex}
\begin{equation}
sel ({\omega_\sigma}) = \frac{|f_{\omega_\sigma}(D)|}{|D|}, \quad  \text{with   } 0 \leq sel ({\omega_\sigma})  \leq 1.
\end{equation}
\vspace{-3ex}
\end{definition}

\begin{definition}
\label{def:join_sel} 
\emph{Join selectivity}:
The selectivity $sel (\omega_\Join)$ of a \emph{windowed join} operator that considers tuples from windows $W_{d_1}$ and $W_{d_2}$ over two input streams $d_1$ and $d_2$ is the ratio of qualifying join partners to the cartesian product for all tuples in the input windows: 
\vspace{-2ex}
\begin{equation}
\label{eq:join_sel}
sel (\omega_\Join) = \frac{|W_{d_1} \Join W_{d_2}|} {|W_{d_1} |\times |W_{d_2} |}, \quad  \text{with   }  0 \leq sel (\omega_\Join)  \leq 1.
\end{equation}
\vspace{-2.5ex}
\end{definition}

\begin{definition}
\label{def:agg_sel} 
\emph{Aggregation selectivity}:
The selectivity $sel (\omega_\xi)$ of a \emph{windowed aggregation} operator that considers tuples in a window $W$ from an input stream $D$, is the ratio of distinct group-by values in the window over the window length: 
\vspace{-1ex}
\begin{equation}
\label{eq:agg_sel}
sel (\omega_\xi) = \frac{|\text{\emph{group-by }}(W_D)|} {|W_D|}, \quad  \text{with   }  0 \leq sel (\omega_\xi)  \leq 1.
\end{equation}
\vspace{-3.5ex}
\end{definition}

In general, selectivities, as well as other DCs (e.g., tuple width) can be derived during the training phase since queries anyways are executed to collect observed metrics (throughput and latency) as training data. However, in the inference phase, this is obviously a challenge since for some optimizations tasks (e.g., the initial placement of operators), a query has not been executed and the DCs are unknown.
Yet the cost model can clearly be used at runtime to re-optimize a query in case it does not meet the desired QoS metrics since then DCs are known from the execution before. 
We think that this approach is justified for DSPS, as queries are anyways long-running. 
In future, we aim to work on learned approaches such as \cite{hilprecht2020deepdb} that can be used to predict DCs for DSPS before execution and thus allow our model also to be used for optimizations before a query starts (e.g., an initial placement decision). 

\vspace{-1ex} 
\paragraph{Hardware-related features}
Hardware characteristics have also a profound impact on the performance of the query in a DSPS, and thereby also on tasks such as placement decisions.
In our featurization, we thus include common properties of hardware such as CPU cores, RAM and disk size. These hardware properties are clustered as \feature{instance size} (small, medium and large), similar to the categorization of major cloud providers. 
These features are also encoded in the graph node to describe the placement characteristic of the operator. 
Clearly, more properties such as network bandwidth can be added in future, which is out of scope for this paper. 

\vspace{-1ex}\subsection{Training and Inference Procedure} \label{subsec:learning_arch}
We train the zero-shot models in a supervised way using Graph Neural Networks (GNNs) to learn from the transferable features.
All categorical transferable features are encoded using a so-called one-hot encoding per operator, while numerical features are normalized. 

During training, the encoded transferable features are used as input to the nodes of the GNN. 
Particularly, the features of each node are encoded by a corresponding node-specific multi-later perceptron (MLP) into a fixed vector (i.e., the hidden state). 
These hidden states of nodes are then combined along the graph, using a bottom-up message passing phase by adding up the hidden vectors from the children nodes and combining them with the parent node of the graph. 
Lastly, the hidden state of the parent node is fed into a final MLP that predicts cost metrics at inference time. 
In the training phase, instead, the predictions are compared to the real costs and the MLPs are updated accordingly.
\begin{table}
  \scriptsize
 \begin{tabular}{ll}
\hline
\textbf{Feature} & \textbf{Training data range} \\ \hline
\texttt{instance size} & small, medium, large \\ \hline
\texttt{input event rate} & {[}0.25, 0.75, 0.5, 1, 1.5, 2.5{]} $\times 10^3$ e/s \\
\texttt{tuple data type} & {[}1...5{]} $\times$  [\texttt{int, string, double}]  \\ \hline
\texttt{filter function} & \textless{},\textgreater{},\textless{}=, \textgreater{}=, !=, \texttt{startswith, endswith} \\
\texttt{literal data type} & \texttt{int, string, double} \\ \hline
\texttt{window type} & sliding, tumbling \\ 
\texttt{window policy} & count-based, time-based \\
\texttt{window size} & {[}0.25, 0.5, 1, 2, 3{]} sec; {[}3, 5, 10, 25, 50, 75, 100{]} tuples \\ 
\texttt{slide size} & {[}0.3 ... 0.7{]} $\times$ window length \\ \hline
\texttt{join-key data type} &\texttt{int, string, double} \\ \hline
\texttt{agg. function} & \texttt{min, max, mean, avg} \\ 
\texttt{group-by data type} & \texttt{int, string, double, none} \\
\hline
\end{tabular}
\caption{Operator-related features for training data generation using synthetic data.}
\label{tab:training_data}
\vspace{-7.5ex}
\end{table}

\vspace{-1.5ex}
\section{Experimental Evaluation}
\label{sec:evaluation}
In this section, we discuss the results of our experimental evaluation. 
We first explain the generation of data that is used for training and the evaluation setup. Afterwards, we demonstrate the accuracy and robustness of our cost model by carrying out various experiments. 

\vspace{-2ex}\subsection{Data Generation and Setup}
\paragraph{Training Data Generation:} 
We created a synthetic training data set with randomly generated queries by enumerating over \textit{(i)} operator parameters, \textit{(ii)} different data streams and \textit{(iii)} distinct query structures. 
Additionally, we evaluated our model against the existing streaming benchmarks~\cite{bordin2020}. 
The different operator parameters and streaming workload configurations (e.g. \feature{event rate}, \feature{data types}) are described in Table \ref{tab:training_data}. 
A variety of queries are evaluated by the use of three query structures as depicted in Figure \ref{fig:queries}. They are composed of widely used streaming operators that we instantiated with different parameters. As a result, a high number of possible combinations of operator properties, streaming workloads and query structures are covered in the training set.

\begin{figure}[t]
\begin{minipage}[p]{0.58\linewidth}
    \includegraphics[width=\linewidth]{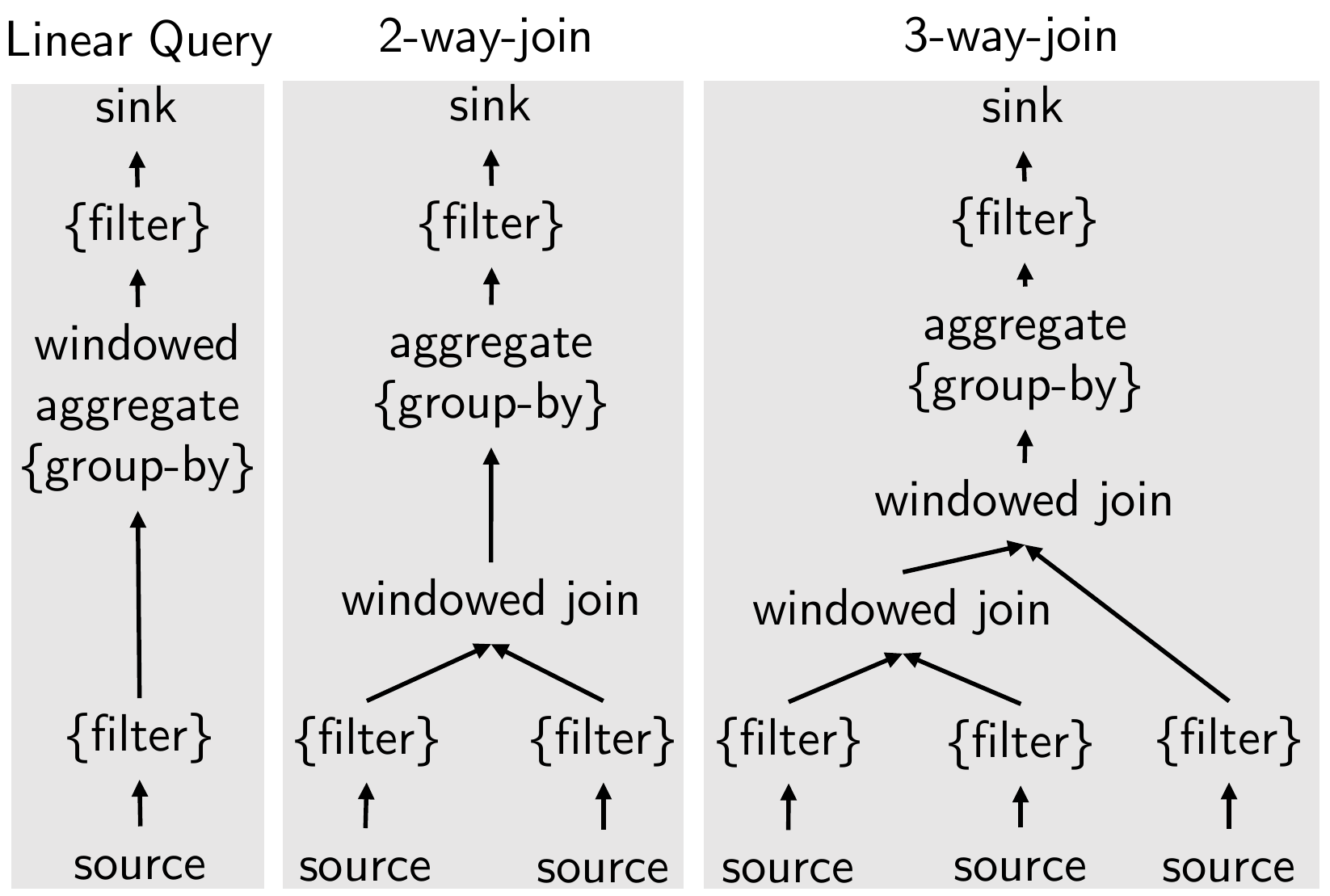}
    \vspace{-5ex}
    \captionof{figure}{Basic query structures used for training data generation. \emph{Filter} operators and \emph{group-by} are optional.}
    \label{fig:queries}
    \vspace{-4ex}
\end{minipage}
\hspace{1ex}
\begin{minipage}[p]{0.38\linewidth}
    \footnotesize
    \begin{tabular}{p{0.6cm}p{0.4cm}p{0.4cm}p{0.4cm}}
    \hline
    \textbf{VM size} 
    & \textbf{CPU} 
    & \textbf{RAM (GB)} 
    & \textbf{Disk (GB)} \\ \hline
    small & 1 & 1 & 10 \\
    medium & 2 & 2 & 20 \\
    large & 8 & 8 & 80 \\ \hline
    \end{tabular}
    \captionof{table}{Hardware-related parameters of virtual instances used for training.}
    \label{tab:instance_sizes}
    \vspace{-5ex}
\end{minipage}
\end{figure}

\vspace{-1ex}\paragraph{Setup and Metrics}
To obtain sufficient training data, 15.000 training queries (5000 per query structure) were generated and executed for 1.5 min over Storm clusters with ten virtual machines of different sizes (small, medium, and large --- S, M, L). 
We used the \textit{Stream-API} of Apache Storm and the standard round-robin scheduler for placement. The worker nodes have properties as specified in Table \ref{tab:instance_sizes}, which is used as \feature{instance size} in the zero-shot model training. 
For queries that did not receive an output tuple at the sink during their runtime due to unfavorable random operator properties, no costs could be determined. 
Consequently, 11.175 queries remained that were divided into training (80\%), validation (10\%) and test (10\%) sets. The model was trained with the training data as explained in \Cref{subsec:learning_arch}.
We report the \textit{Q-error} metric $q(c, \hat{c})$ to describe the relative deviation of the true cost metric value $c$ (latency or throughput) and its prediction $\hat{c}$, which is a standard metric for the quality of cost models, defined as: $q(c, \hat{c}) = max \left(\frac{c}{\hat{c}}, \frac{\hat{c}}{c} \right)$, with $q \geq 1$ \cite{hilprecht2022}. 
\vspace{-2ex}
\subsection{Experimental Results}
In the following evaluation, we investigate the accuracy of the zero-shot cost models on different workloads. 
In the experiments, we mainly focus on the generalization ability to predict \first for an unseen test set but using parameters inside the defined training range, \ii for workloads with parameters outside the defined training range, \iii for unseen workloads from existing streaming benchmarks and \iv for workloads under varying hardware heterogeneity.

\vspace{-1ex}\paragraph{Q1: How accurate is the model for an unseen combination of operator and workload properties (inside the training range)?}
At first, we provide the observed Q-errors in predictions for the test set. These are queries and data streams not considered during the training, but use values from the same ranges as shown in Table
\ref{tab:training_data}.
It can be seen in in Table \ref{tab:test_set} that the median and the $95$th percentile of the Q-error of our model is highly accurate for the given query structures.
Typically, cost models based on heuristics have much higher Q-errors of up to $300$ for the $95$th percentile or more \cite{hilprecht2020deepdb}.

\begin{table}
\scriptsize
\begin{tabular}{lllll}
\hline
\textbf{Query structure} & \multicolumn{2}{l}{\textbf{Latency}} & \multicolumn{2}{l}{\textbf{Throughput}} \\
 & \multicolumn{1}{l}{median} & 95th & \multicolumn{1}{l}{median} & 95th \\ \hline
Linear query & 1.09 & 2.14 & 1.15 & 2.68 \\
2-way-join & 1.13 & 2.69 & 1.14 & 2.91 \\
3-way-join & 1.21 & 5.25 & 1.22 & 5.59 \\ \hline
Overall & 1.13 & 3.19 & 1.16 & 3.50 \\ \hline
\end{tabular}
\caption{Q-errors (median and 95th percentile) for unseen combinations of operator and workload properties inside the training range. } 
\vspace{-10.5ex}
\label{tab:test_set}
\end{table}

\vspace{-1ex}\paragraph{Q2: How accurate is the model for completely unseen workloads (outside the training range)?} 
To answer this, we generated queries that are within the known parameter space except for one property that exceeds or falls below the training value range. 
For each property, we generated 50 queries (for all three query structures). 
The median Q-error from all estimations is reported in \Cref{fig:extrapolation}. 
To be more precise, we tested tuple widths larger than the training set (\textsf{A}) as well as lower and higher event rates (\textsf{B}). 
To investigate extrapolation for unseen operator properties, we applied bigger time-based (\textsf{C}) and count-based (\textsf{D}) window sizes as well. 
In almost all cases, a fairly low median Q-error could be achieved. 
Intuitively, the Q-error increases with the distance to the training value range as more extreme properties will lead to more extreme costs, which are hard to predict precisely; still way better than existing heuristics~\cite{hilprecht2020deepdb}.
In a second set of experiments, we observed the accuracy for unseen query structures; i.e., we use 4-way and 5-way-joins instead of only up to 3-way joins that were used for training. Moreover, for linear queries, we added 2 to 4 additional filter operators after the \feature{windowed aggregate} operator (cf. \Cref{tab:query_structures}). 
While the median Q-errors for those are also promising, the tail accuracy of the unseen query structures increases with the increasing query complexity. 
We aim to tackle this in a follow-up work by incorporating extreme unseen properties in an additional fine-tuning phase (called \emph{few-shot learning}), where only a few unseen queries are used for retraining.

\begin{figure}
\includegraphics[width=\linewidth]{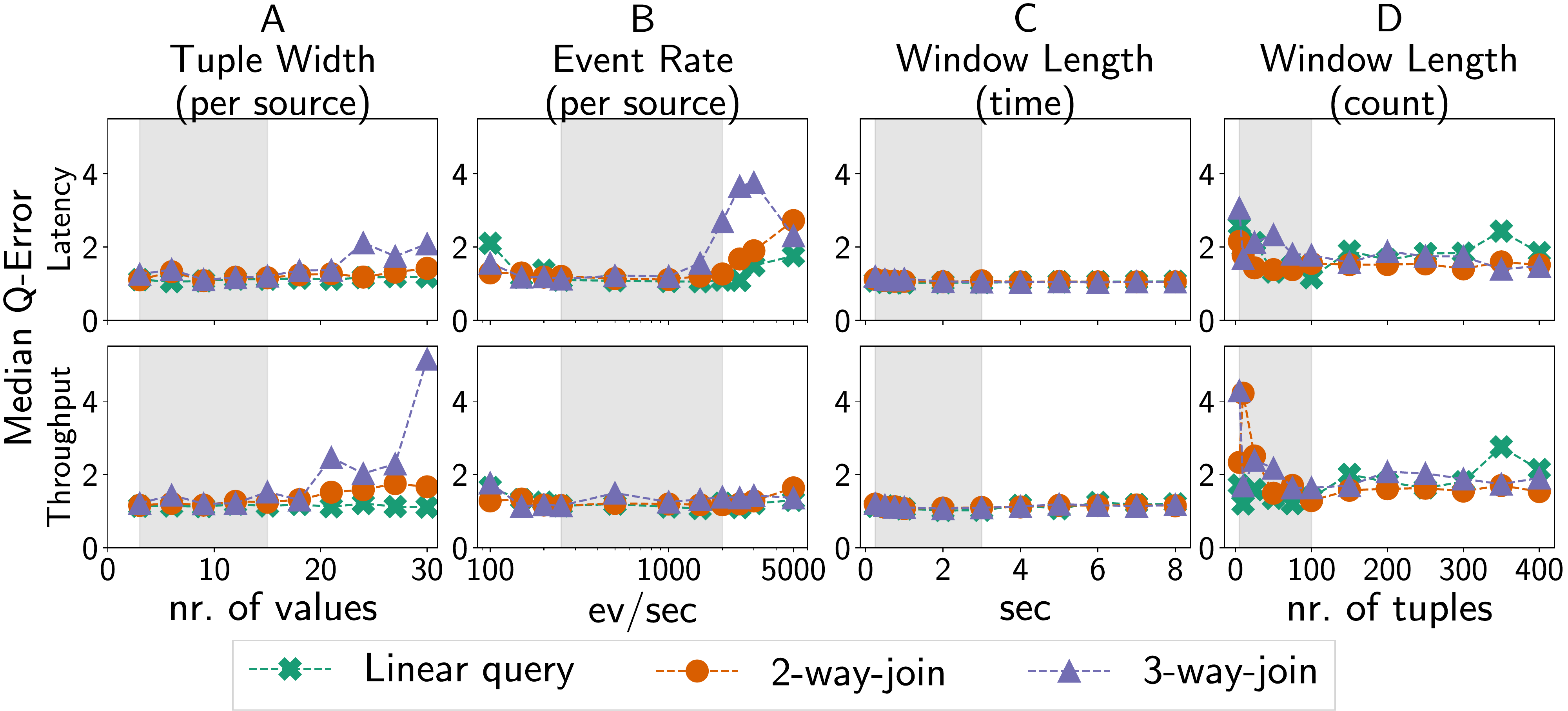}
\vspace{-5ex}
\caption{Median Q-errors for queries with unseen tuple widths (\textsf{A}), event rates (\textsf{B}), time-based and count-based window lengths (\textsf{C, D}).  The gray shaded area marks the training range. The model extrapolates for unseen queries and workloads (white shaded areas), often with very high accuracy.}
\label{fig:extrapolation}
\vspace{-2ex}
\end{figure}

\begin{table}
\scriptsize
\begin{tabular}{lllll}
\hline
\textbf{Query structure} & \multicolumn{2}{l}{\textbf{Latency}} & \multicolumn{2}{l}{\textbf{Throughput}} \\
 & \multicolumn{1}{l}{median} & 95th & \multicolumn{1}{l}{median} & 95th \\ \hline
2-filter-chain & 1.14 &2.41 &1.59&3.65 \\
3-filter-chain & 2.67 &46.34&2.82&27.78 \\
4-filter-chain & 7.33 &54.68 &3.94&59.73 \\ \hline
4-way-join & 1.95 &24.30 &1.33 &20.79 \\
5-way-join & 1.91 &26.76 &1.35& 21.87\\\hline
\end{tabular}
\caption{Q-errors for unseen query structures. Although the model has never seen these query structures during training, it is able to make accurate cost predictions for them.}
\label{tab:query_structures}
\vspace{-9ex}
\end{table}

\vspace{-1ex}\paragraph{Q3: How accurate is the model for existing benchmark workloads?} 
To evaluate the generalization capability of the model, we applied it on a selection of three different existing benchmark workloads \cite{bordin2020}.
\first The \emph{Advertisement Analysis} is a benchmark that computes the ratio of aggregates of two incoming streams (i.e. clicks, impressions), grouped by two attributes. 
In our evaluation, we used two sub-queries with a windowed aggregation for both of these streams and a third sub-query that joins both streams. 
\ii The \emph{Spike Detection} benchmark is derived from an IoT use case that computes and compares the moving average of sensor values.
\iii The \emph{Smart Grid} benchmark targets the computation of energy consumption in households from the DEBS Grand Challenge 2014. 
Its \emph{outlier detection} task is decomposed into three sub-queries as presented in \cite{koliousis2016}. 
We consider two sub-queries that contain a large sliding window of 60 min (and 60s sliding length) and build average over the local and the global energy consumption. 
As our model was trained with window lengths up to 3 secs, we reduce the benchmark window length to 5 min (and 5s sliding length), which is still is $100\times$ higher than our training range.
As no event rates were specified in the benchmark, we execute the queries five times with different rates each and report the results in Table \ref{tab:benchmark}.
This shows that the zero-shot model can adapt to streaming workloads of existing benchmark out-of-the-box and estimates the costs very accurately.

\begin{table}
\scriptsize
\begin{tabular}{lllll}
\hline
\textbf{Benchmark} & \multicolumn{2}{l}{\textbf{Latency}} & \multicolumn{2}{l}{\textbf{Throughput}} \\
 & \multicolumn{1}{l}{median} & 95th & \multicolumn{1}{l}{median} & 95th \\ \hline
 
Advertisement (clicks) & 1.51 & 1.53 & 1.38 & 1.39 \\
Advertisement (impressions) & 1.51 & 1.52 & 1.38 & 1.39 \\
Advertisement (join) & 1.99 & 2.06 & 1.55 & 2.16 \\
Spike Detection & 1.01 & 1.04 & 1.73 & 1.94 \\
Smart Grid (local) & 1.21 & 1.23 & 1.92 & 1.92\\
Smart Grid (global) & 1.20  & 1.66 & 1.91 & 1.91 \\ \hline
\end{tabular}
\caption{Q-errors for benchmarks from \cite{bordin2020}. Each query has been executed 5 times with different event rates. This shows our model is able to accurately predict out-of-the-box for these queries.}
\label{tab:benchmark}
\vspace{-11ex}
\end{table}

\vspace{-1ex}\paragraph{Q4: Does the model predict appropriately for different hardware types?} 
As streaming operators are placed on distinct instances (cf. \Cref{tab:instance_sizes}), we expect the model to take the underlying node performance into account during the prediction. 
To investigate that, we take all queries from the test set and obtain the predictions under the assumption that all operators are either located on a small, medium or large instance and report the mean of the predictions. 
Given the linear query, for instance, the model predicts for larger instances higher throughput (31.17/36.14/39.98 ev/sec for S/M/L, resp.).
Hence, the zero-shot model correctly learns the effect of instance size on the query performance. In further studies, we aim to model and evaluate hardware characteristics more fine-grained.
\vspace{-4.5ex}
\section{Related Work}
\label{sec:related_work}
We classify the existing work on cost estimation into two main categories: \first analytical approaches, heuristics and \ii learning-based approaches as stated below. 

\vspace{-1.2ex}\paragraph{Analytical Approaches and Heuristics} 
Several analytical approaches have been applied to predict resource costs in DSPS, such as using queuing theory \cite{mencagli2018, dematteis2017} for different DSPS optimization tasks. 
However, analytical approaches often make assumptions about the data distribution of incoming streaming workload that renders the estimates either inaccurate or ill-suited for DSPS due to the inherently dynamic nature of the streaming workload. 
Many works propose heuristics to optimize tasks like placement, namely, greedy approaches~\cite{aniello2013}, meta heuristics~\cite{chandramouli2011} and custom heuristics~\cite{eskandari2021}.
Storm~\cite{toshnival2014} also adopts heuristics for placement~\cite{nardelli2019}.
However, heuristics often also make simplifying assumptions leading to sub-optimal decisions for the optimization tasks. 
Flink DSPS~\cite{carbone2015} uses a query optimizer from Stratosphere \cite{alexandrov2014}, however, it explicitly targets only user-defined operators in the cost modelling. 

\vspace{-1.2ex}\paragraph{Learned Stream Processing}
Approaches related to resource cost estimation are the closest to this work. 
Regression models have been used extensively for estimating 
performance metrics such as latency \cite{li2018} and throughput \cite{imai2017, wang2017} in DSPS. 
However, these works only consider a limited set of features and largely ignores featurization related to queries that have a major impact on the performance. 
Different learning methods have been applied to optimize DSPS tasks~\cite{ahmad2021}, namely for elasticity \cite{russo2019} or operator placement \cite{li2018, luthra2021}.  
Similar to this work, \cite{lombardi2018} employs neural networks to model the performance of DSPS; however, they stumble around the challenge of re-training for a large number of heterogeneous queries and workloads. 
In contrast, we introduce a data and query agnostic approach that can support a variety of stream processing optimization tasks that are dependent on the cost estimation of resources \emph{without} any cost of additional training.

\vspace{-1.5ex}
\section{Conclusion}
\label{sec:conclusion}
In this paper, we show that zero-shot cost models are highly effective in predicting costs even in the presence of unseen workload conditions. 
The model can robustly predict QoS metrics for a wide range of queries out of a large set of possible configurations. 
As such, we believe that the cost estimations are readily usable for optimization tasks of DSPS, such as the operator placement.
In the future, an interesting direction is to solve optimization tasks like placement \emph{out-of-the-box} and more explicitly modelling hardware and network properties in the proposed graph structure. 

\vspace{0.5ex}
\small{
\textbf{Acknowledgments.}
This work has been supported by the IPF program and the Cloud Computing Competence Centre of DHBW Mannheim, the Collaborative Research Center (CRC) 1053 MAKI funded by the German Research Foundation (DFG), the NHR4CES Program, hessian.AI at TU Darmstadt as well as DFKI Darmstadt. 
}

\vspace{-1ex}
\balance{}
\bibliographystyle{IEEEtran}
\bibliography{bibliography.bib}

\end{document}